\documentclass[%
 aip,
 amsmath,amssymb,
reprint,%
]{revtex4-1}
\usepackage{color}
\usepackage{graphicx}% Include figure files
\usepackage{dcolumn}% Align table columns on decimal point
\usepackage{bm}% bold math
\newcommand{\kcb}[1]{\textcolor{black}{#1}}
\usepackage[utf8]{inputenc}
\usepackage[T1]{fontenc}
\usepackage{mathptmx}

\begin{document}

\preprint{AIP/123-QED}

\title{High frequency guided mode resonances in mass-loaded, thin film gallium nitride surface acoustic wave devices}
% Force line breaks with \\

\author{Stefano Valle}
 \affiliation{Quantum Engineering Technology Labs and Department of Electrical and Electronic Engineering, University of Bristol, Woodland Road, Bristol BS8 1UB, UK}%Lines break automatically or can be forced with \\
\author{Manikant Singh}%
\affiliation{Center for Device Thermography and Reliability, H.H. Wills Physics Laboratory, University of Bristol, Woodland Road, Bristol BS8 1UB, UK}
\author{Martin Cryan}
\affiliation{Quantum Engineering Technology Labs and Department of Electrical and Electronic Engineering, University of Bristol, Woodland Road, Bristol BS8 1UB, UK}
\author{Martin Kuball}%
\affiliation{Center for Device Thermography and Reliability, H.H. Wills Physics Laboratory, University of Bristol, Woodland Road, Bristol BS8 1UB, UK}
\author{Krishna C. Balram}
 \email{krishna.coimbatorebalram@bristol.ac.uk}
\affiliation{Quantum Engineering Technology Labs and Department of Electrical and Electronic Engineering, University of Bristol, Woodland Road, Bristol BS8 1UB, UK}
    
\date{\today}% It is always \today, today,

\begin{abstract}
We demonstrate high-frequency (> 3 GHz), high quality factor radio frequency (RF) resonators in unreleased thin film gallium nitride (GaN) on sapphire and silicon carbide substrates by exploiting acoustic guided mode \kcb{(Lamb wave)} resonances. The associated energy trapping, due to mass loading from the gold electrodes, allows us to efficiently excite these resonances from a 50 $\Omega$ \kcb{input. The higher phase velocity, combined with lower electrode damping, enables high quality factors with moderate electrode pitch, and provides a viable route towards  high-frequency piezoelectric devices.} The GaN platform, with its ability to guide and localize high-frequency sound on the surface of a chip \kcb{with access to high-performance active devices,} will serve as a key building block for monolithically integrated RF front-ends.  
\end{abstract}

\maketitle

%\section{\label{sec:Intro}Introduction:}

A modern smartphone has $\sim$50 RF filters to enable seamless communication across a wide variety of frequency bands \cite{ruby2015snapshot}. As we move towards widespread adoption of 5G standards, and expect ever-greater functionality from our phones, the number of RF filters is expected to significantly increase and the filter packaging problem becomes increasingly acute \cite{lam2016review}. Monolithic integration of piezoelectric acoustic wave filters with amplifiers is the only scalable long-term solution. While integration of aluminum nitride (AlN) film bulk acoustic wave resonator (FBAR) filters with complementary metal-oxide semiconductor (CMOS) electronics is being actively pursued, the process incompatibility between micro-electro-mechanical systems (MEMS) and CMOS foundries, mainly substrate release, makes monolithic integration challenging. In the past decade, three trends have converged to make GaN an interesting alternative \cite{ansari2012monolithic,rais2014gallium}: improvements in material growth, especially interface quality for thin films, steady displacement of GaAs based power-amplifiers (PA) with GaN PAs and finally, the availability of GaN foundry services providing ready-access to high performance amplifiers. \kcb{ Working with GaN allows us to trade-off a lower piezoelectric device  performance (cf. AlN, lithium tantalate) for the benefits of monolithic integration with active devices}. 

The prospect of building piezoelectric, primarily surface acoustic wave (SAW), resonators and filters in GaN was recognized from the beginning of GaN device research \cite{lee2001epitaxially} and there has been a lot of exciting recent progress \cite{neculoiu2018band}. On the other hand, relatively little work has been done on exploiting the main advantage that GaN provides over traditional SAW substrates like lithium tantalate \cite{muller2015sezawa}. In particular, GaN supports guided acoustic waves whose dispersion can be engineered, on account of its lower acoustic velocity, compared to the growth substrate (mainly silicon, silicon carbide and sapphire) and buffer layers \cite{fu2019phononic}. Guiding sound on the surface of the chip allows us to confine acoustic energy and increase the effective electromechanical coupling coefficient \cite{takai2017ihp}. In addition, by trapping these sound waves using guided mode resonances\cite{fan2002analysis}, one can engineer high quality factor ($Q_{mech}$) resonators without requiring substrate release. Such unreleased resonator platforms are also being explored for monolithic integration in non-piezoelectric CMOS-MEMS platforms \cite{wang2011acoustic,marathe2012si,bahr2015theory,erbes2019acoustic}. In this work, we demonstrate high $Q_{mech}$ GaN resonators on unreleased substrates (sapphire and silicon carbide) by utilizing the metallic interdigitated transducer (IDT) electrodes for mass loading and energy trapping\cite{shockley1967trapped}. By avoiding patterning of the GaN device layer, our approach is compatible with current GaN HEMT foundry process flows and is attractive for the near-term monolithic integration of piezoelectric acoustic wave devices with GaN HEMT amplifiers.
%\section{Device design and operation}
\begin{figure}[hbtp]
    \centering
    \includegraphics[width = \linewidth]{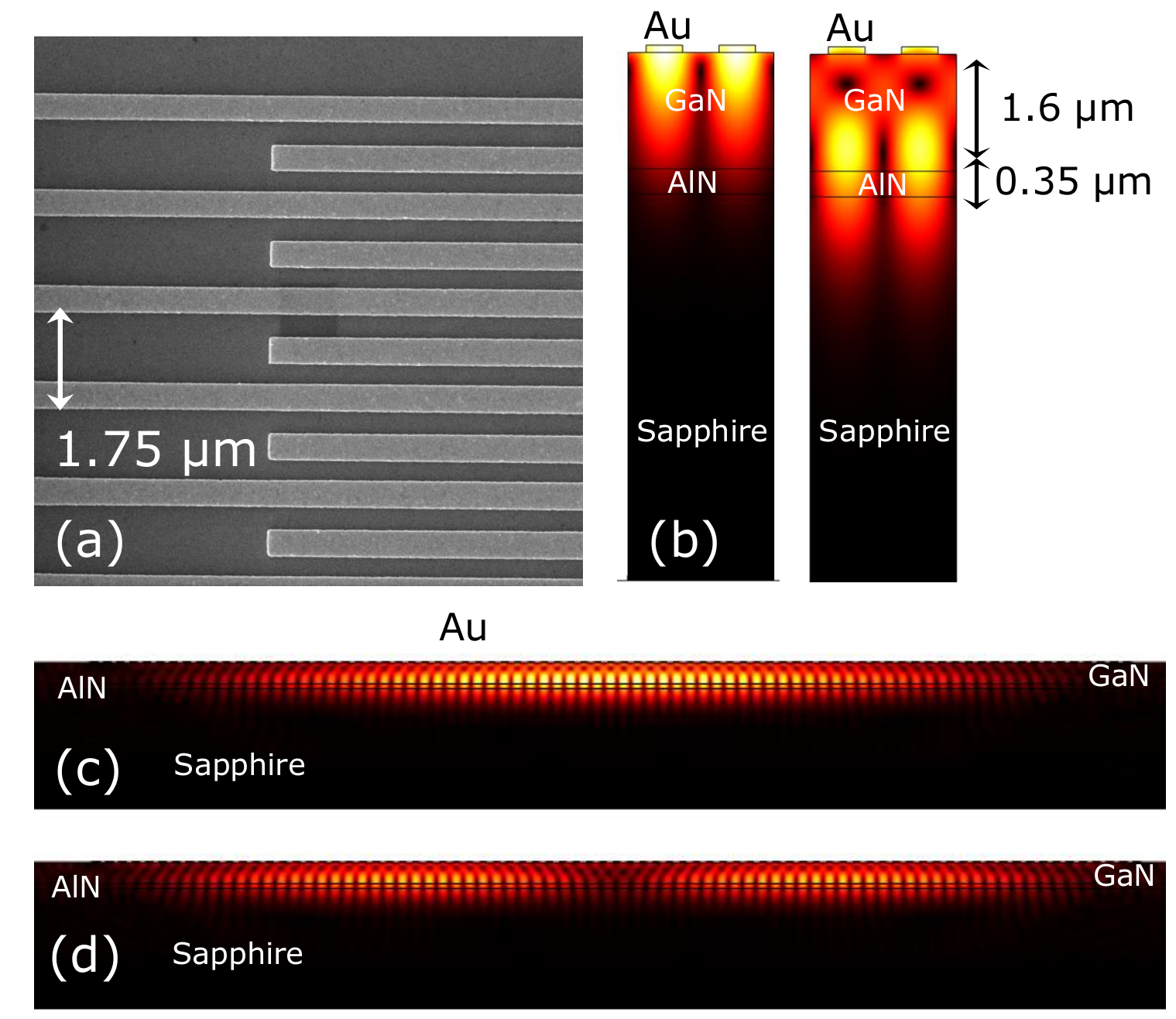}
    \caption{(a) SEM image of a fabricated GaN-on-SiC device (${\Lambda}_{IDT}$ = 1.75 ${\mu}m$, $t_{Au}$ = 75 nm) (b) Finite element method (FEM) simulation \kcb{performed using linear elastic piezoelectric theory  in COMSOL Multiphysics v5.3} of the mode displacement \kcb{of a 1.6 ${\mu}m$ GaN on sapphire device with 350 nm AlN buffer layer (${\Lambda}_{IDT}$ = 1.75 ${\mu}m$, $t_{Au}$ = 75 nm)} for the SAW ($f_{SAW}\approx$ 2.06 GHz) and Lamb wave ($f_{Lamb}\approx$ 3.02 GHz) modes (c) FEM simulation of the mode displacement for the fundamental \kcb{($f\approx$ 3.056 GHz)} and higher-order \kcb{($f\approx$ 3.049 GHz)} trapped Lamb wave resonances ($N_{pairs}$ = 40).}
    \label{fig:1_schematic}
\end{figure}

A representative device, shown in Fig.\ref{fig:1_schematic} (a), consists of the standard IDT geometry, with the IDT period (${\Lambda}_{IDT}$) chosen to match the SAW frequency. The devices were fabricated either on SiC or sapphire substrates with the GaN thickness ($t_{GaN}\approx$ 1-1.5 ${\mu}m$) and electrode thickness ($t_{Au}\approx$ 75 nm).  In addition to exciting the Rayleigh SAW waves (shown in Fig.\ref{fig:1_schematic} (b, left)), the transducer with the same period (${\Lambda}_{IDT}$), also excites a higher-frequency guided Lamb wave mode (\kcb{also referred to as a Sezawa mode}), shown in Fig.\ref{fig:1_schematic} (b, right) \cite{takagaki2002guided,camou2001guided,muller2015sezawa}. \kcb{In contrast to the Rayleigh SAW mode, the Lamb wave mode has a higher phase velocity ($\sim$1.5x) allowing higher frequency operation for the same $\Lambda_{IDT}$. In addition, a significant fraction of the acoustic energy resides within the GaN layer, away from the GaN-electrode interface, leading to lower acoustic dissipation and higher $Q_{mech}$ at high-frequencies (2-6 GHz). }

 To convert the IDT from a SAW transducer to a SAW resonator, one needs to confine the acoustic energy, which can be achieved by employing acoustic reflectors on either side of the IDT. Traditionally, SAW resonators have relied on the  weak finger reflectivity provided by shorted metallic gratings. But the weak reflectivity increases the number of finger pairs needed to achieve unity reflection, resulting in large device sizes ($\sim mm$). In contrast, we work in the regime of strong finger reflection ($r_{f}\approx 0.06-0.1$) \cite{hashimoto2004design} and excitation of trapped acoustic modes to achieve compact high-$Q_{mech}$ resonators. While higher finger reflectivity is generally associated with higher out-of-plane scattering, the scattered waves phased appropriately can destructively interfere in the far-field (substrate leakage), leading to high quality factors and tight mode confinement. Photonic and phononic crystal structures, similarly rely on strong index contrast \cite{joannopoulos2008molding} to simultaneously achieve high $Q$ and low mode volumes. In addition, the increased radiation conductance associated with energy trapping allows us to match the device impedance to 50 $\Omega$ and avoid the need for on-chip impedance matching networks \cite{datta1986surface,sarabalis2019role}. Effectively, the increased radiation conductance ($G_{IDT}$) dominates the IDT static capacitance ($C_{0,IDT}$) and thus resistive matching is sufficient \cite{datta1986surface}.
 
The higher $r_{f}$ is achieved by using gold (Au) electrodes for the IDT. For the same thickness, a Au electrode was measured to have a higher ($\approx$ $50x$) $r_{f}$ for SAW waves compared to aluminum (Al) electrodes \cite{soluch2014effect}. In a guided wave geometry, the $r_{f}$ is higher on account of the higher spatial mode overlap of the acoustic field with the metal electrode. Traditionally, Au electrodes have not been used with SAW devices, especially at high-frequencies ($>1$ GHz), due to the high acoustic wave damping. \kcb{Compared with traditional Rayleigh wave-based SAW devices that have acoustic energy confined to the electrode-GaN interface and thus suffer from high damping, the Lamb wave has a significant fraction of the acoustic energy below the interface (Fig.\ref{fig:1_schematic}(b)) and thus reduced spatial mode overlap with the electrode. Coupled with the higher phase velocity, which increases the IDT period and reduces the transducer series resistance, Lamb waves provide a viable alternative to traditional SAW and FBAR devices for building high-frequency, high-$Q_{mech}$ piezoelectric devices.}

If the $r_{f}$ exceeds a threshold value determined by the total number of fingers in the structure, the structure resembles a 1D acoustic Bragg stack with the IDT electrodes (GaN+Au) and the inter-electrode gaps (GaN) serving as the two effective acoustic materials comprising the stack. The strong finger reflection couples the forward and backward propagating Lamb waves, leading to a 1D phononic bandgap. Since the guided modes in the IDT region have a lower frequency (due to mass loading) compared to the bare GaN regions, the overall structure can be visualized as a 1D Bragg stack surrounded by two semi-infinite bare GaN regions that do not support propagating modes at the same frequency. Provided the frequency difference ($\Delta\omega$) for the guided mode in the IDT and non-IDT regions exceeds the modal decay rate ($\kappa$), the mode is trapped in the IDT, and a guided mode resonance can be observed in an RF reflection ($|S_{11}|$) measurement. We would like to note here that the trapped acoustic modes are analogous to the bound states observed in GaAs/AlGaAs superlattice structures and a similar transfer-matrix based tunneling resonance approach can be used to calculate them\cite{ko1988matrix}. Using an FEM simulation, the trapped Lamb wave resonances' displacement profiles can be calculated and two of these are shown in Fig.\ref{fig:1_schematic} (c,d). The fundamental mode has the highest quality factor and electromechanical coupling coefficient, with the higher-order modes being more weakly confined. While we have performed experiments on both GaN-on-SiC and GaN-on-sapphire devices, we restrict our FEM simulations to GaN-on-sapphire.

\begin{figure}[htbp]
    \centering
    \includegraphics[width = \linewidth]{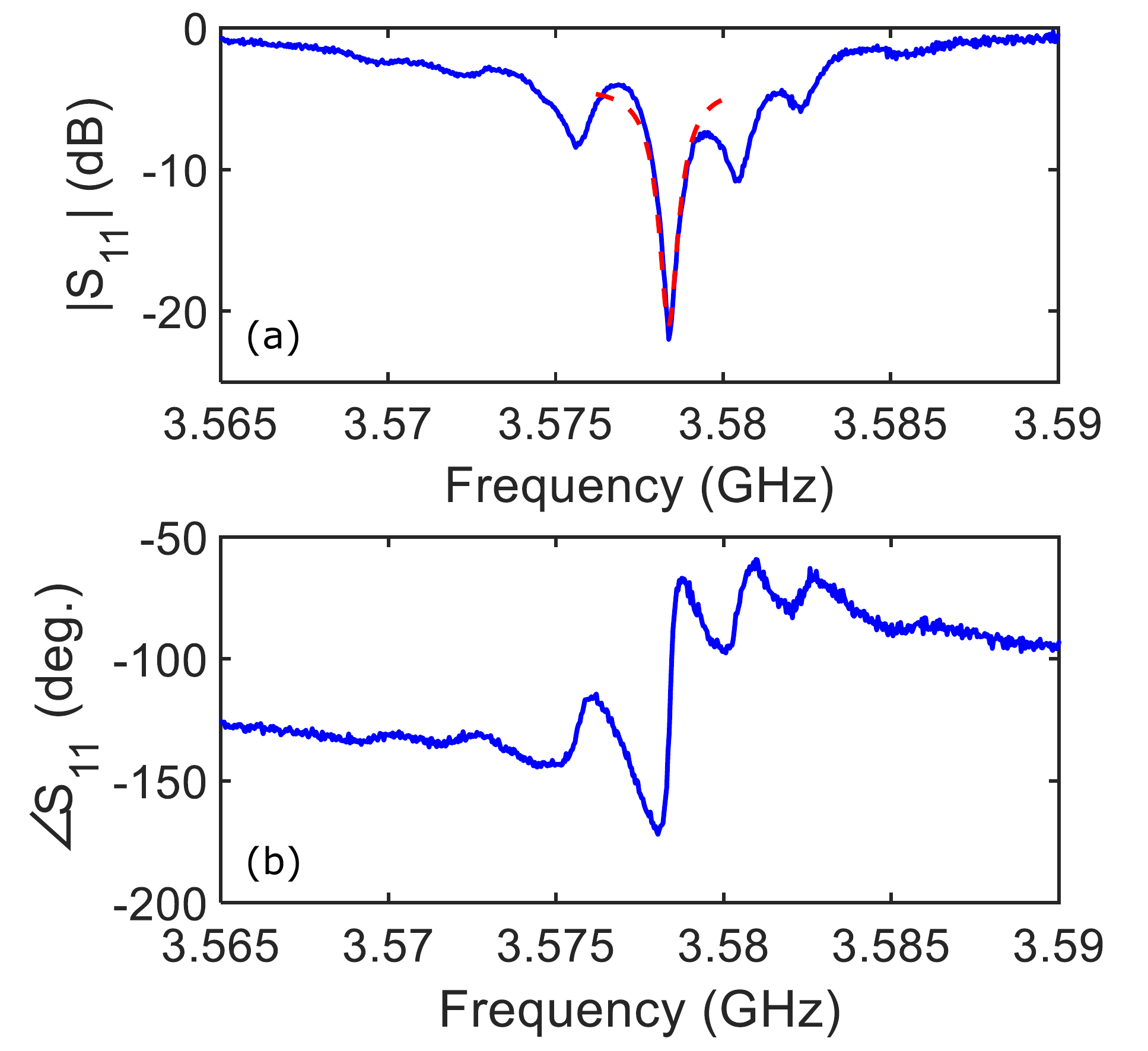}
    \caption{(a) Measured RF reflection coefficient magnitude $|S_{11}|$ and (b) phase response for a GaN on SiC device with 125 finger pairs and IDT period 1.5 ${\mu}$m. The red dashed curve in (a) shows a Lorentzian fit to the fundamental mode response.}
    \label{fig:2_S11_carbide}
\end{figure}

These guided mode resonances can be observed in the RF reflection ($|S_{11}|$) spectrum of the IDT. A representative spectrum, for a 1.5 ${\mu}$m IDT period device fabricated on a GaN on SiC sample, is shown in Fig.\ref{fig:2_S11_carbide}. In contrast to the expected $sinc^2(f)$ response for the $S_{11}$ magnitude as in a traditional IDT, we instead observe a series of sharp Lorentzian resonances, corresponding to the different trapped modes. The effect can also be observed in the phase response, where a series of phase inflections can be clearly seen in Fig.\ref{fig:2_S11_carbide}(b),  corresponding to the dips in Fig.\ref{fig:2_S11_carbide}(a). The fundamental resonance, fit with a Lorentzian lineshape (red curve in Fig.\ref{fig:2_S11_carbide}(a)) has a quality factor , $Q_{mech}\approx$ \kcb{2000}, which corresponds to an $f.Q\kcb{{\sim}7*10^{12}}$. Since acoustic waves at such high frequencies are expected to be strongly scattered by GaN growth defects, such as threading dislocations, such high $Q_{mech}$ point to the high-material quality in a GaN-SiC platform, especially at the GaN-buffer interface.  \par

%\section{Energy trapping and dependence of resonator $Q_{mech}$ on device parameters}
\begin{figure}[htbp]
    \centering
    \includegraphics[width = \linewidth]{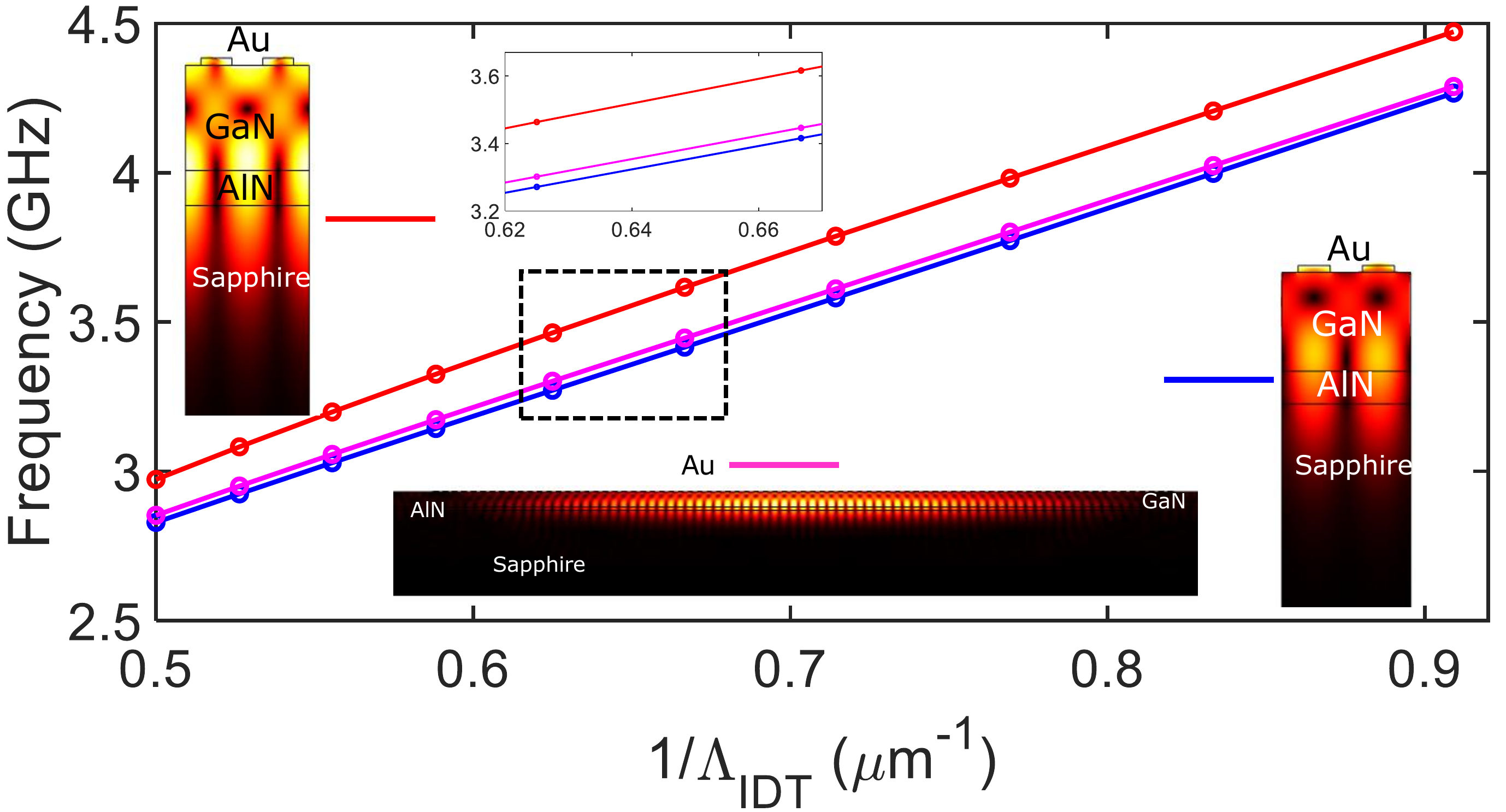}
    \caption{Simulated frequencies of the two non-degenerate higher order Lamb wave modes (blue, \kcb{inset right} and red, \kcb{inset left}) as a function of inverse electrode period \kcb{$1/{\Lambda}_{IDT}$}. The \kcb{respective} mode displacement profiles \kcb{for the two modes} are shown in the inset. The \kcb{Lamb wave resonance} (inset, bottom) frequency ($N_{pairs}$ = 40) is also shown (magenta). A zoomed-in section (shown by the dashed box) of the plot is shown in the inset for visualization. Simulation parameters: $t_{GaN}$ = 1.5 ${\mu}m$, $t_{AlN}$ = 0.35 ${\mu}m$ and $t_{Au}$ = 75 nm}
    \label{fig:freq_splitting}
\end{figure}

To quantify the effects of $r_{f}$, we can use the idea of degenerate mode-splitting. A bare GaN on sapphire substrate unit cell supports two degenerate Lamb wave modes (forward and backward propagating). Adding a gold electrode (as part of the IDT) couples the forward and backward modes, lifting the frequency degeneracy and forming two modes (shown as a function of IDT period, by the red and blue curves in Fig.\ref{fig:freq_splitting}). The frequency splitting between the two modes is governed by the coupling between the original forward and backward propagating modes, and thus is a measure of $r_{f}$. For the IDT unit-cell, Fig.\ref{fig:freq_splitting} is plotting the 1D phononic bandgap at the $\Gamma$ point as a function of $\Lambda_{IDT}$. It is important to keep in mind that only the lower mode is excited by the IDT due to RF excitation symmetry. Also shown in Fig.\ref{fig:freq_splitting} is the trapped fundamental mode frequency ($N_{pairs}$ = 40) and we can see that the trapped modes lie within the 1D phononic bandgap (zoomed inset in Fig.\ref{fig:freq_splitting}). As long as the frequency separation ($\Delta\omega\approx$ 30 MHz) of the bound mode (\kcb{magenta}) from the guided band (blue) is greater than the mode decay rate ($\kappa\approx$ 3-5 MHz in our devices), a sharp Lorentzian response, corresponding to the guided mode resonance, will be observed in the RF reflection spectrum and the IDT acts as a resonator. On the other hand, if $\Delta\omega\sim\kappa$, then bound mode spectrum overlaps with the guided mode band and the IDT acts as a transducer.

\begin{figure}[htbp]
    \centering
    \includegraphics[width = \linewidth]{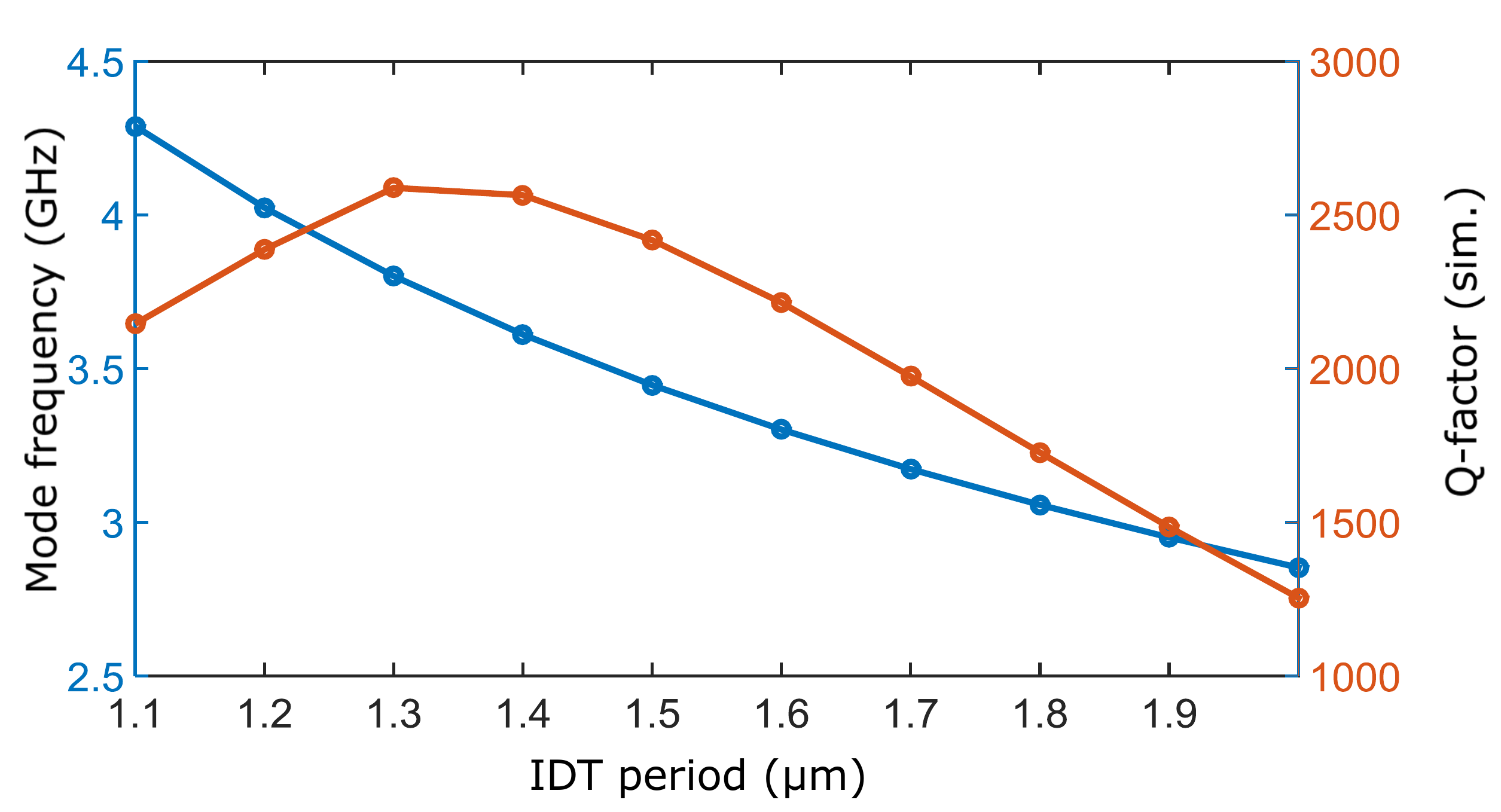}
    \caption{Simulated Lamb wave fundamental resonance mode frequency (blue) and quality factor (brown) for GaN-on-sapphire devices with varying IDT periods ($N_{pairs}=40$). Simulation parameters: $t_{GaN}$ = 1.5 ${\mu}m$, $t_{AlN}$ = 0.35 ${\mu}m$ and $t_{Au}$ = 75 nm.}
    \label{fig:Q_period}
\end{figure}
From a resonator design perspective, it is critical to understand the dependence of center frequency ($f_{c}$) of the resonator on the GaN device layer thickness ($t_{GaN}$), as that becomes the key fabrication parameter to control. While the IDT-based design does ensure that the $f_{c}$ changes with IDT period, the quality factor ($Q_{mech}$) of the guided mode resonance is not constant as a function of frequency. Given a GaN layer thickness, there is an optimal frequency range ($f_{opt}$) for achieving high $Q_{mech}$. At $f{\ll}f_{opt}$, the finger reflection drops due to reduced $t_{Au}/\lambda_{acoustic}$. At the other extreme $f{\gg}f_{opt}$, the out-of-plane scattering and metal damping increases due to increased finger reflection and greater spatial mode overlap with the metal, on top of the standard $f^2$ acoustic wave attenuation dependence\cite{auld1973acoustic}. We can visualize the confinement dependence on IDT period by plotting the simulated quality factor ($Q_{sim}$) for the fundamental Lamb wave resonance, for devices with N = 40 finger pairs ($t_{GaN}$ = 1.5${\mu}m$, $t_{AlN}$ = 0.35${\mu}m$ and $t_{Au}$ = 75 nm). The results are shown in Fig.\ref{fig:Q_period}.  The quality factor ($Q_{sim}$) is calculated from an eigenfrequency calculation by $Q_{sim}=f_{real}/2f_{imag}$, \cite{comsol} where $f_{real}$ and $f_{imag}$ correspond to the real and imaginary components of the complex eigenfrequency. It is important to note that the FEM calculation does not properly account for high-frequency metal damping and the $Q_{sim}$ is primarily an estimate of out-of-plane (substrate) scattering losses. As the plot shows, there is an optimal period for a given combination of $t_{GaN}$ and $t_{Au}$ to achieve the highest Lamb wave $Q_{mech}$. From the plots, we can construct a rule-of-thumb of $\approx 500$ MHz of $f_{c}$ tuning using the IDT period, while maintaining a high $Q_{sim} > 2000$.   

%\section{From multimode to single mode resonators}

\begin{figure}[htbp]
    \centering
    \includegraphics[width = \linewidth]{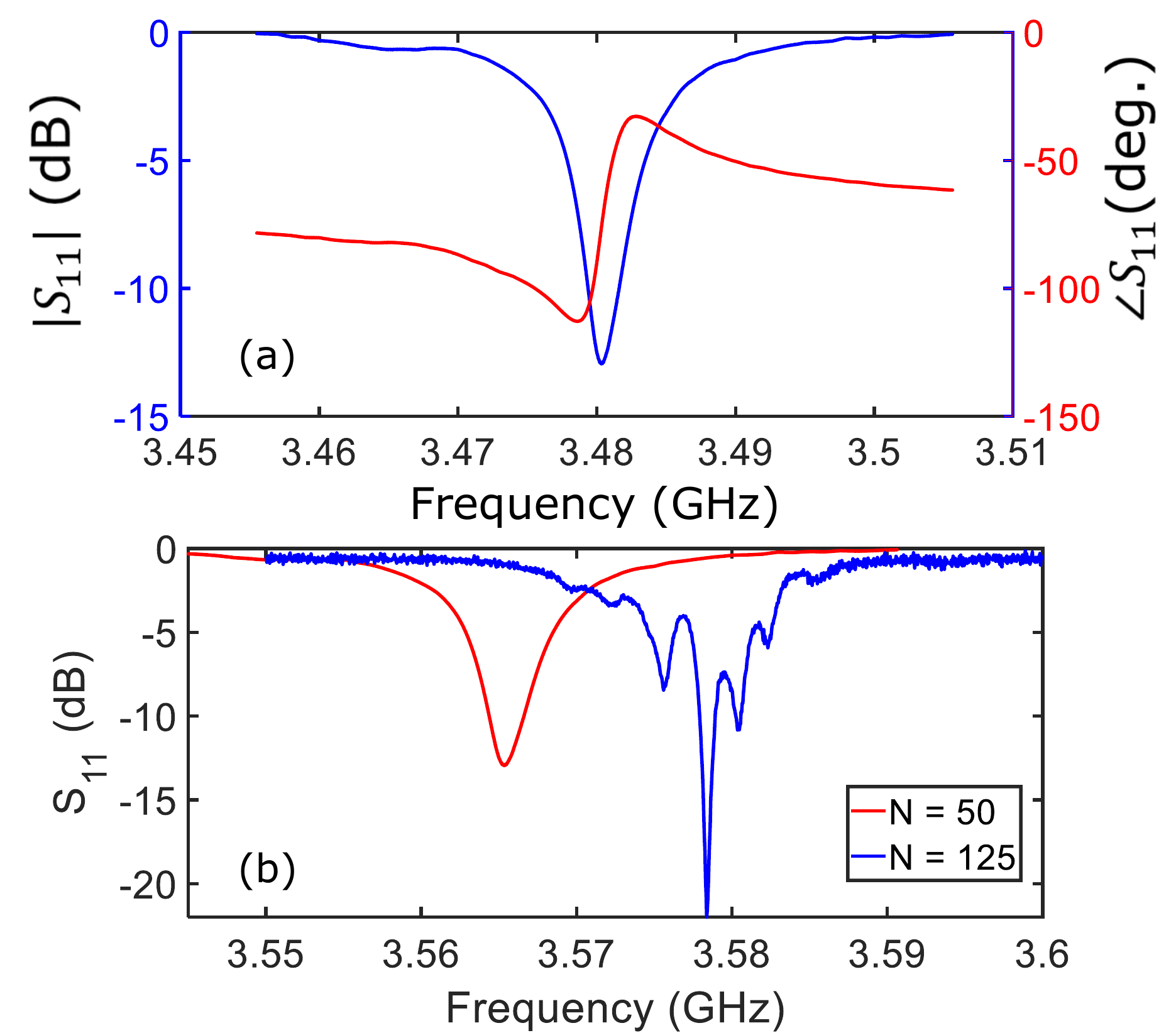}
    \caption{(a) Measured RF reflection coefficient magnitude $|S_{11}|$ and phase $\angle{S_{11}}$ response for a GaN on SiC device with 50 finger pairs and IDT period 1.75 ${\mu}$m. (b) Comparison of the measured RF $|S_{11}|$ response for the 50 ($\Lambda_{IDT}$ = 1.75 ${\mu}m$) and 125 ($\Lambda_{IDT}$ = 1.5${\mu}m$) finger pairs devices. The response of the N=50 device has been shifted in frequency for easier comparison.}
    \label{fig:3_SM_MM}
\end{figure}

While the $Q_{mech}$ achieved by the device in Fig.\ref{fig:2_S11_carbide} is high, the presence of multiple closely spaced resonances distorts both the amplitude and phase response, making it challenging to use in practical applications. Ideally, one would like to work with a single isolated resonance with high $Q_{mech}$ so that more complex elements, like coupled resonator filters, can be synthesized. The IDT device in Fig.\ref{fig:2_S11_carbide} had 125 finger pairs and as shown in Fig.\ref{fig:1_schematic} (c,d), in a uniform period (duty-cycle) IDT, adjacent sections of the transducer can support resonant modes, as shown in Fig.\ref{fig:1_schematic}(d), leading to an inherent multi-moded response.  By reducing the number of finger pairs, we can effectively achieve single-mode operation in the device, shown in Fig.\ref{fig:3_SM_MM}. As both the amplitude ($|S_{11}|$) and phase ($\angle{S_{11}}$)  response show in Fig.\ref{fig:3_SM_MM}(a), the response is primarily dominated by the fundamental mode. On the other hand, reducing the cavity size leads to lower mode confinement and reduced mode quality factors ($Q_{mech}$). This can be clearly seen in Fig.\ref{fig:3_SM_MM}(b) where the multi-mode and single mode (frequency-shifted) responses are plotted on the same scale. The single-mode device has a $Q_{mech}\approx 500$, compared to the $Q_{mech}\approx \kcb{2000}$ for the multi-mode device.  Ideally, one would like to combine the high $Q_{mech}$ achievable with large N, with a single-mode response to enable design of more complicated circuit elements like coupled resonator filters.

\begin{figure}[htbp]
    \centering
    \includegraphics[width = \linewidth]{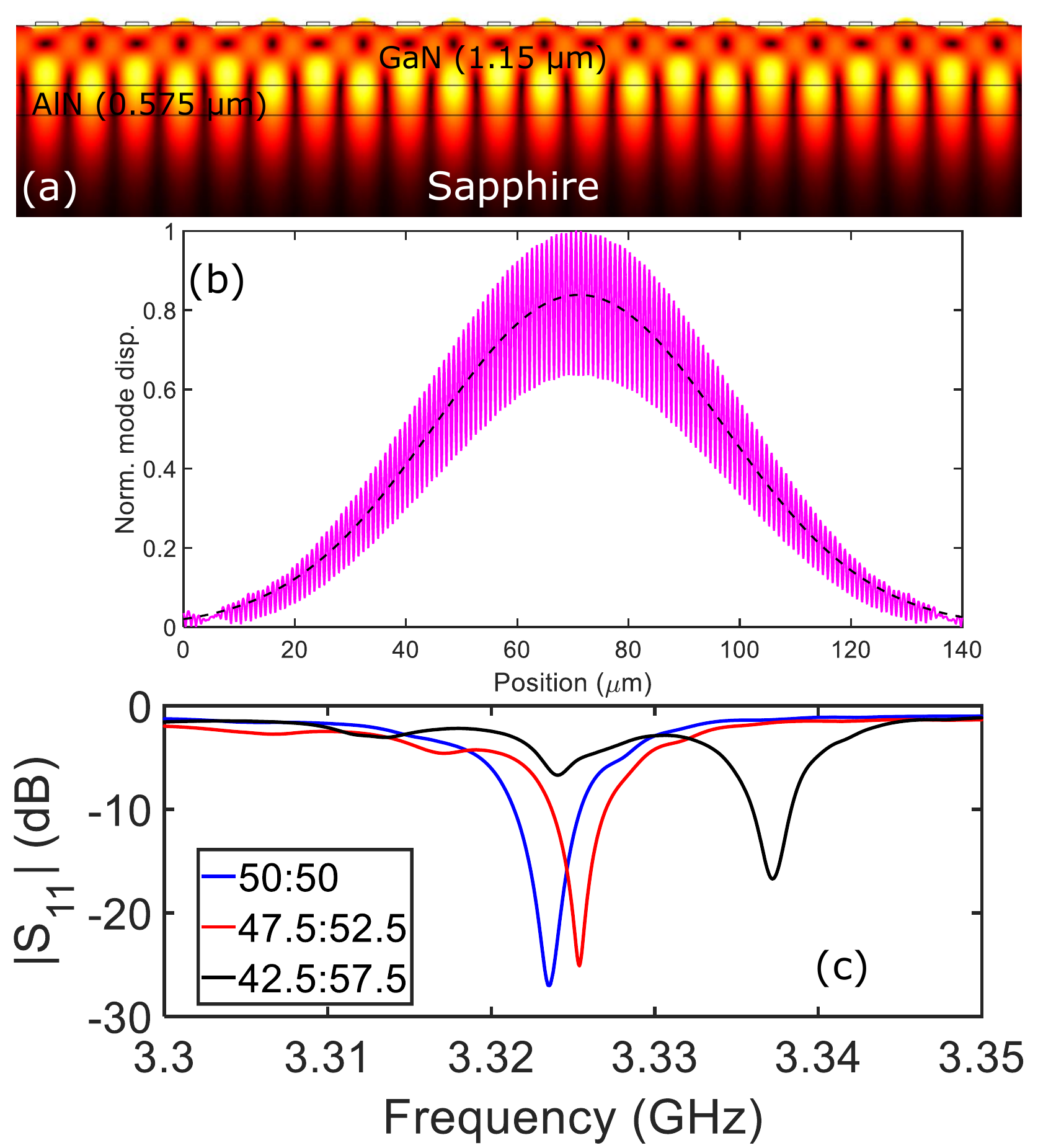}
    \caption{(a) Mode displacement of the trapped Lamb wave mode in a GaN-on-sapphire device (b) Line-cut of the mode displacement (magenta) alongwith a Gaussian fit (black) (c) Measured RF reflection ($|S_{11}|$) response of apodized IDTs fabricated on a GaN-sapphire platform ($t_{GaN} = 1.1\kcb{5} {\mu}m$, \kcb{$t_{AlN}$ = 0.575 ${\mu}m$}, $t_{Au} = 75 nm$) with different duty cycle chirps. \kcb{The $Q_{mech}$ of the Lamb wave resonance of each device, calculated from a Lorentzian function fit to the $S_{11}$ power spectrum, are $\sim$ 322, 386  and 478 respectively}}
    \label{fig:Phononic_Xtals}
\end{figure}

%\section{Improving $Q_{mech}$: towards 1D phononic crystal designs}
To implement a high-$Q_{mech}$ single-mode response, we can borrow ideas from the photonic crystal community on Gaussian 1D confinement \cite{quan2011deterministic,tanaka2008design} of electromagnetic fields to achieve extremely high quality factors. In contrast to the optical domain where the refractive index is shaped in response to plane wave excitation, in the acoustic case, the metal electrodes provide both the excitation field and the confinement potential (due to the mass loading) and thus the mode shape and excitation efficiency are intimately linked, constraining the designs. Fig.\ref{fig:Phononic_Xtals} (a) shows the mode shape for a uniform period IDT and Fig.\ref{fig:Phononic_Xtals} (b) plots the mode displacement as a function of position. As the red Gaussian curve fit shows, the fundamental cavity mode shape can be well-approximated by a Gaussian with a peak displacement in the cavity centre.

By apodising the duty cycle of the IDT to match the Gaussian mode-shape, we can both improve the $Q_{mech}$ and achieve single-mode operation. The improvement in $Q_{mech}$ occurs due to two effects: since the peak of the modal displacement occurs in the cavity centre, by reducing the duty cycle of the IDT, we are effectively reducing the overlap of the acoustic field with the metal, which is the major source of damping and scattering. By apodizing the grating, we also reduce the $\vec{k}$-space overlap between the \kcb{Lamb wave resonance} and the substrate modes, reducing the overall out-of-plane leakage \cite{srinivasan2002momentum}. Apodising the \kcb{IDT-period} also ensures single mode operation as the local duty cycle chirp ensures that higher order modes (such as those shown in Fig.\ref{fig:1_schematic}(d)) are not trapped. Fig.\ref{fig:Phononic_Xtals}(c) shows the measured RF $|S_{11}|$ spectrum of three devices with identical period and number of finger-pairs ($N_{pairs}$ = 100), but varying duty-cycle Gaussian chirp fabricated on a 1\kcb{.15} ${\mu}m$ GaN on Sapphire platform. The measurement results indicate that as we increase the duty-cycle chirp, \kcb{we observe increased mode separation, a key step towards single-mode operation}. We also see that the excitation efficiency (peak $|S_{11}|$ dip) is not uniform and \kcb{attribute} this to the mode excitation and confinement being provided by the same electrodes. We are currently working on achieving independent control of the excitation and confinement by patterning the GaN layer\cite{xu2018high}. \kcb{Although the phononic crystal designs do show an improvement in $Q_{mech}$, the GaN-on-sapphire devices do not achieve the performance of the uniform $\Lambda_{IDT}$ GaN-on-SiC devices, primarily due to the inferior quality of the GaN-substrate interface, which leads to excess acoustic scattering and dissipation.}

\begin{figure}[htbp]
    \centering
    \includegraphics[width = \linewidth]{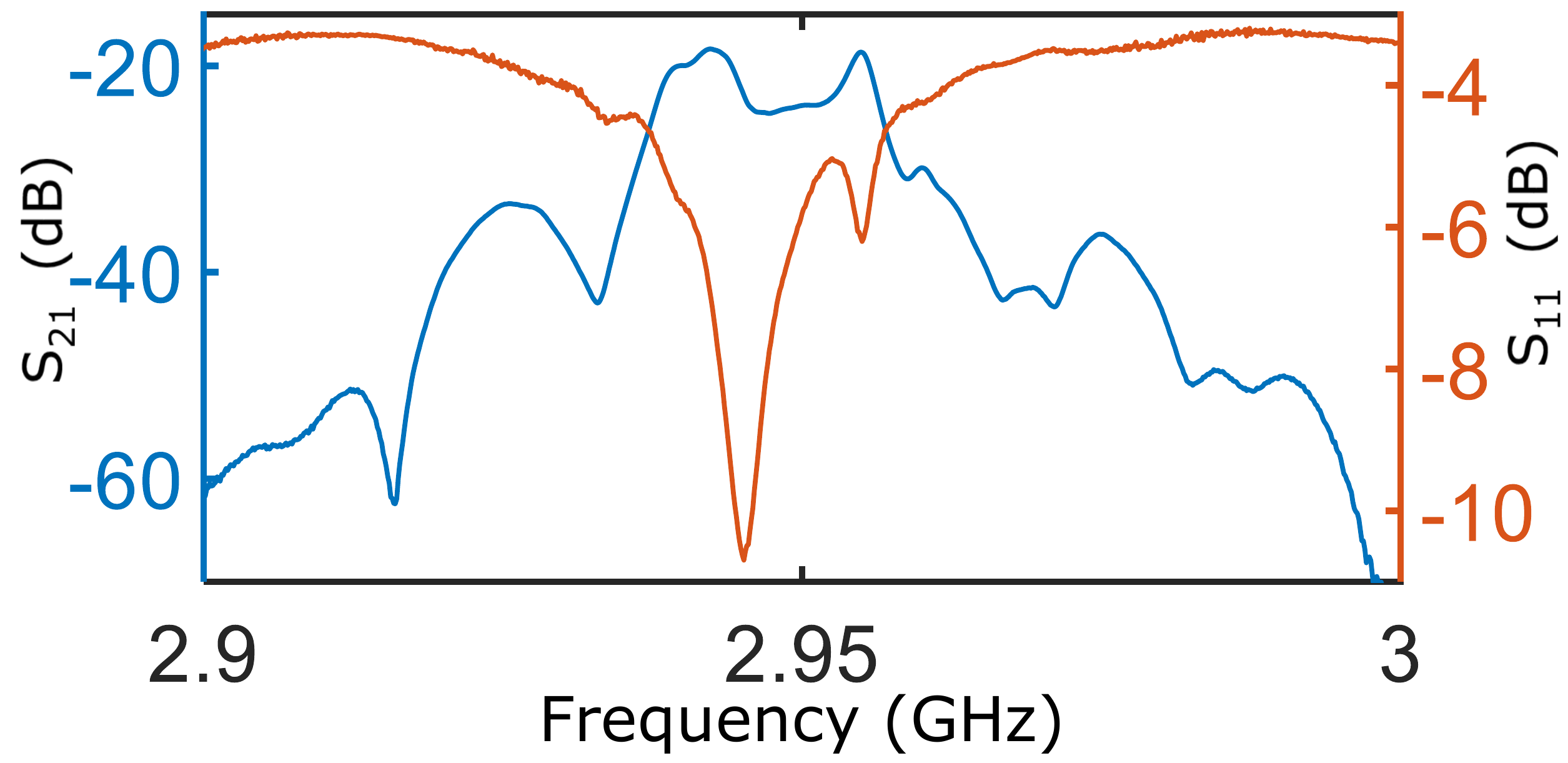}
    \caption{\kcb{RF transmission ($S_{21}$) spectrum of a coupled resonator device fabricated on a GaN on SiC platform ($\Lambda_{IDT}=$ 2 ${\mu}m$ and resonator gap = 10 ${\mu}m$) . The resonator response ($S_{11}$) is overlaid for reference.}}
    \label{fig:S21_data}
\end{figure}

\kcb{Finally, as a first step towards integrated resonant Lamb wave filters in GaN, Fig.\ref{fig:S21_data} shows the RF transmission spectrum for a coupled resonator device with $\Lambda_{IDT}$ = 2 ${\mu}m$ and resonator spacing 10 ${\mu}m$. The transmission spectrum is peaked around the Lamb wave resonances, which can be seen in the overlaid $S_{11}$ spectra. We would like to note that while the insertion loss is high ($\sim$ 18 dB), there is tremendous scope for performance improvement by optimising the device geometry (ensuring high-$Q_{mech}$ single mode operation) and filter design (controlling the coupling rate between the two resonators).}\par 

\textbf{Acknowledgements}: K.C.B would like to thank \kcb{M-A. Campana}, V. Gokhale, J. Haine, M. Uren, and E.T-T. Yen for valuable discussions and suggestions. We would like to acknowledge funding support from the European Research Council (SBS3-5 758843) and Engineering and Physical Sciences Research Council (EP/N015126/1).
 
%\section{References}

\bibliography{APL_refs}

\end{document}